\documentclass{JHEP3}

\usepackage{amsmath}
\usepackage{epsfig}
\usepackage{graphicx}

\newcommand{\be}{\begin{equation}}
\newcommand{\ee}{\end{equation}}
\newcommand{\bea}{\begin{eqnarray}}
\newcommand{\eea}{\end{eqnarray}}
\newcommand{\ba}{\begin{array}}
\newcommand{\ea}{\end{array}}
\newcommand{\nn}{\nonumber
\\}

\font\mybb=msbm10 at 10pt
\def\bb#1{\hbox{\mybb#1}}

\def\bE {\bb{E}}

\makeatletter 
\@addtoreset{equation}{section} 
\makeatother 
 
\def\appendix#1{
  \addtocounter{section}{1}
  \setcounter{equation}{0}
  \renewcommand{\thesection}{\Alph{section}}
  \section*{Appendix \thesection\protect\indent \parbox[t]{11.15cm}
  {#1} }
  \addcontentsline{toc}{section}{Appendix \thesection\ \ \ #1}
  }

\preprint{UB-ECM-PF-08/07\\
DAMTP-2008-20}

\title{Accelerating Branes and Brane Temperature}

\author{J. G. Russo\\
Instituci\'o Catalana de Recerca i Estudis
Avan\c cats (ICREA),
\\
Departament ECM and Institut de Ciencies del Cosmos, 
\\
Facultat de F\'{\i}sica, Universitat de Barcelona,
\\
Diagonal 647, E-08028 Barcelona,
Spain.\\
}

\author{P. K. Townsend\\
Department of Applied
Mathematics and Theoretical Physics \\
Centre for Mathematical
Sciences, University of Cambridge\\
Wilberforce Road, Cambridge, CB3
0WA,
UK.\\ 
}

\abstract{We define the local acceleration and jerk of a relativistic brane in an ambient spacetime, and construct from them a dimensionless parameter  $\lambda$ that must be small for an interpretation of  brane acceleration as  local Unruh temperature.  As examples, we discuss (i) open rotating branes, for which 
$\lambda>1$ (ii) closed spherical  branes  expanding in Minkowski  spacetime, for which 
$\lambda=0$ when the worldvolume is either an Einstein static universe or de Sitter space, in which case the brane temperature equals the Gibbons-Hawking temperature, (iii) closed spherical branes  in anti-de Sitter spacetime, for which a maximally symmetric worldvolume is  anti-de Sitter, Minkowski or de Sitter according to whether the magnitude of the brane acceleration is  less than, equal to or greater than a `critical' value, and (iv)  the BTZ black hole, viewed as a membrane. }


\begin{document}

\section{Introduction}
\setcounter{equation}{0}

Following Hawking's discovery that asymptotically-flat stationary black holes emit thermal radiation \cite{Hawking:1974sw},  Unruh showed that there is a similar effect in flat space for a particle detector undergoing constant  proper acceleration $A$: the detector appears to be in a heat bath at the Unruh temperature \cite{Unruh:1976db} (see \cite{Crispino:2007eb} for a recent review)
\be\label{Unruh}
T = \frac{A}{2\pi}\, . 
\ee
As Unruh pointed out, the two phenomena are related. Near  the black hole event horizon, the metric is approximately the direct product of a  2-dimensional Rindler spacetime with the round  2-sphere metric, 
and an application of the Unruh formula to the Rindler spacetime yields the Hawking temperature 
$T_H$ of the black hole blue-shifted according to the Tolman law for thermodynamic equilibrium in stationary spacetimes with timelike Killing vector field $\xi$:
\be\label{Tolman}
T= T_H/\sqrt{-\xi^2}\, .
\ee
A more direct connection between the Hawking and Unruh temperatures was established by Deser and Levin \cite{Deser:1998bb,Deser:1998xb}. They considered the global embedding of the Schwarzschild metric in a six-dimensional Minkowski spacetime \cite{Fronsdal:1959uu}.
A static observer,  at  fixed distance from the horizon,  undergoes constant acceleration in the ambient flat spacetime  such  that the Unruh temperature equals the local temperature of the observer. This is possible because the acceleration has both a component  tangent to the spacetime {\it and} a component orthogonal to it; near the horizon the acceleration is tangential but at infinity it is 
orthogonal.  These results have since been extended to other black holes \cite{Chen:2004qw,Chen:2004ky,Santos:2004ws,Hong:2000kn,thorla}.

The idea of an Unruh-type interpretation of local temperature via a global embedding in a higher-dimensional spacetime first arose in the context of the Gibbons-Hawking (GH) temperature 
of de Sitter (dS)  space \cite{Gibbons:1977mu} (see also \cite{Figari:1975km}),  which can be viewed as a hypersurface in a Minkowski spacetime of one higher dimension. In that case, an application of Unruh's formula yields  \cite{Narnhofer:1996zk}
\be\label{thirring}
2\pi\, T= A = \sqrt{a^2 + R^{-2}}\, , 
\ee
where $a$ is the magnitude of the acceleration in de Sitter space and $R$ is the de Sitter radius. 
One recovers the GH temperature by considering the class of observers with $a=0$. 
The general application of this idea is often referred to as  the GEMS  (Global Embedding in Minkowski Spacetime) program, but  it may be necessary to  consider higher-dimensional flat spacetimes that are {\it not} Minkowski\footnote{We are not aware of any theorem that guarantees the existence of a global isometric embedding of a generic pseudo-Riemannian metric in a flat spacetime (in contrast to the Riemannian case) although local embeddings of this type always exist \cite{Friedman:1961}.}. A simple example is  $d$-dimensional anti-de Sitter space, which is a hypersurface in $\bE^{(2,d-1)}$;  using this embedding, one can show  that an observer undergoing a constant acceleration $a$ in the anti-de Sitter space has an acceleration $A$ in the flat embedding space such that
\be\label{acceladS}
A^2 = a^2 - R^{-2}\, . 
\ee
This is negative if $a<1/R$, which is possible when the embedding spacetime has two time dimensions. In such cases, Unruh's result applies only to spacelike $A$, so we deduce that the temperature experienced by an observer undergoing constant acceleration $a\ge 1/R$ in anti de Sitter space of radius $R$ is given by \cite{Deser:1997ri}
\be\label{DL}
2\pi\, T= \sqrt{ a^2 - R^{-2}}\, . 
\ee
The temperature experienced by an observer undergoing constant acceleration of magnitude $a\le 1/R$ is zero \cite{Bros:2001tk}. 

The main aim of this paper is to apply these ideas to charged relativistic branes by considering the motion of a `probe' $p$-brane in a $D$-dimensional Minkowski, de Sitter or anti de Sitter spacetime. In the latter two cases we may view the motion in the context of the flat $(D+1)$-dimensional spacetime in which the (anti) de Sitter spacetime is globally embedded, so in all cases we have a flat `ambient' spacetime.  By `relativistic'  we mean a $p$-brane with energy density equal to its tension $\mu$, and by `charged' we mean that it couples minimally to an external $(p+1)$-form potential $C$ with $(p+2)$-form field strength $F=dC$. Given arbitrary local worldvolume coordinates $\sigma^i$ ($ i=0,1,\dots, p)$, the effective low-energy dynamics is governed by the Dirac-type action
\be\label{Dirac}
S= -\mu \int d^{p+1}\sigma \left[ \sqrt{-\det g} + \,  {\cal C}\right]\, ,
\ee
where $g$ is the induced worldvolume metric, and  ${\cal C}$ is the worldvolume Hodge dual of the worldvolume $(p+1)$-form induced by $C$. In cartesian coordinates $X^\mu $ for the flat ambient spacetime, 
the equation of motion is 
\be\label{eofm}
g^{ij}K_{ij}^\mu + F_{(ext)}^\mu =0\, , \qquad \ \mu=0,...,D-1\ ,
\ee
where $K_{ij}^\mu$ is the extrinsic curvature tensor of the worldvolume, and $F^\mu_{(ext)}$ the `external' force per unit $p$-volume exerted by the external form field:
\be
F^{(ext)}_\mu = \frac{1}{\left(p+1\right)!\sqrt{-\det g}}\, \varepsilon^{i_1\dots i_{p+1}} \partial_{i_1}X^{\nu_1} \cdots \partial_{i_{p+1}}X^{\nu_{p+1}}F_{\mu \nu_1\dots \nu_{p+1}}\, . 
\ee
Observe that $\partial_i X\cdot F_{(ext)} \equiv 0$, as required for consistency since the extrinsic curvature satisfies the `Brane Bianchi' identity 
\be\label{BBI}
K_{ij} \cdot \partial_k X \equiv 0\, . 
\ee

\subsection{Brane acceleration}

An immediate problem for our program is that the brane equation of motion (\ref{eofm}) does not  involve the concept of acceleration in any obvious way. One way around it would be to introduce a congruence of timelike worldvolume worldlines corresponding to the trajectories of `test' particles with worldvolume velocity field $v^i$ and worldvolume acceleration field $a^i(v)=v^j\partial_jv$; recall that $v^2\equiv -1$ and $v_ia^i(v)\equiv0$.  The $D$-velocity in  the ambient flat spacetime of these test particles is then $V=v^i\partial_i X$, and the $D$-acceleration is
\be
A(V) \equiv v^i\partial_i V = K_{vv}+ a^i(v)\partial_i X\, , \qquad K_{vv}\equiv v^iv^jK_{ij} \, .
\ee
As a consequence of the Brane Bianchi identity, it follows that
\be\label{acce}
A^2(v) = K_{vv}^2 +  a^2(v) \, . 
\ee
Now consider a solution of (\ref{eofm}) for which 
\be\label{KadS}
K_{ij}^\mu = - \frac{1}{p+1}\, g_{ij} F_{(ext)}^\mu\, . 
\ee
Recalling that the (intrinsic) Riemann curvature tensor is given in terms of  the extrinsic curvature tensor by the formula
\be\label{GC1}
R_{ijkl} =  K_{ik} \cdot  K_{jl} - K_{il} \cdot K_{jk}\, ,
\ee
we see that the {\it intrinsic} curvature tensor corresponding to  (\ref{KadS}) is 
\be\label{intrinsic}
R_{ijkl} =  {F_{(ext)}^2\over (p+1)^2} \left( g_{ik}  g_{jl} - g_{il}  g_{jk}\right)\, , 
\ee
from which it follows that the worldvolume geometry is conformal to dS if $F_{(ext)}$ is spacelike, and conformal to adS if $F_{(ext)}$ is timelike (which is possible if the ambient spacetime has two time dimensions); it is Minkowski  if $F_{(ext)}$ is null (which allows non-zero  $F_{(ext)}$ if the ambient spacetime has two time dimensions). Using eq. (\ref{KadS}) in  (\ref{acce}) we deduce that
\be
A^2(v) = a^2(v) +  {F_{(ext)}^2\over (p+1)^2} \, .
\ee
For the special case of a constant uniform electric-type field of strength $E$, the force field $F_{(ext)}$ is a fixed $D$-vector of magnitude $|E|$, and hence the world-volume geometry is either de Sitter or anti-de Sitter with radius  $R = (p+1)/|E|$, or Minkowski when  $R=\infty$. The acceleration in such cases is
\be
A^2(v) = a^2(v) \pm R^{-2} \, , \qquad 
\ee
where the plus sign is for de Sitter and the minus sign is for anti de Sitter. An application of the Unruh formula (\ref{Unruh}) for constant $a(v)=a$ yields the results quoted earlier for the temperature of an observer undergoing constant acceleration $a$ in (a)dS. Related observations have been made in \cite{Tian:2005yj} and  \cite{Jennings:2005vv}. 

However,  the introduction of  test particles on the brane is unsatisfactory because the brane does not come supplied with such particles. Moreover, their introduction obscures the extent to which  the results 
are intrinsic to the brane.  We therefore propose a different approach, related to the Hamiltonian formalism that we review in the following section.  In this approach, the worldvolume is assumed to be foliated by spacelike hypersurfaces, with local coordinates $\{\sigma^a; a=1,\dots,p\}$, and the leaves of the foliation are parametrized by some arbitrary worldvolume time $t$. The induced worldvolume metric 
is then written as
\be
g_{ij}d\sigma^id\sigma^j = g_{tt}dt^2 + 2 g_{ta}dtd\sigma^a + h_{ab}d\sigma^ad\sigma^b\, , 
\ee
so that $h_{ab}$ are the components of the induced metric on the brane at fixed time; we denote by
$h^{ab}$ the components of its  inverse. The inverse to this worldvolume metric is
\be\label{inverseg}
g^{ij}\partial_i\partial_j = -u^iu^j\partial_i\partial_j + h^{ab}\partial_a\partial_b
\ee
where the worldvolume one-form dual to $u^i\partial_i$ is 
\be\label{uform}
u_i d\xi^i = -\Delta dt\, , \qquad \Delta \equiv \sqrt{-g_{tt}+ g_{ta}h^{ab}g_{tb}}\, , 
\ee
which means that $u^2=-1$. We interpret $u$ as a worldvolume velocity field, and 
\be\label{UUU}
U^\mu= u^i\partial_i X^\mu 
\ee
as the $D$-velocity of the brane, as defined by the chosen foliation of the worldvolume.  The brane momentum density is then
\be\label{PU}
P^\mu= \mu\,  \sqrt{\det h}\, U^\mu\, , 
\ee
and we define the brane acceleration by
\be\label{acceleration}
A^\mu= u^i\partial_i U^\mu\, . 
\ee
For the special case of $p=0$ these expressions reduce to the standard definitions of  relativistic velocity, momentum and acceleration of a point particle in Minkowski spacetime. 

The expression (\ref{acceleration}) can be rewritten as
\be\label{altA}
A^\mu= K_{uu}^\mu + a^i\partial_i X^\mu\, , \qquad K_{uu}^\mu\equiv u^iu^j K_{ij}^\mu\, , 
\ee
where the worldvolume acceleration is
\be
a^i = u^jD_ju^i \equiv u^j\partial_j u^i + u^ju^k\Gamma_{jk}^i\, . 
\ee 
Observe that $a$ is orthogonal to $u$, as expected, and also that
\be\label{BraneA}
A^2 = K_{uu}^2 + a^2\, ,
\ee
as a consequence of the Brane Bianchi identity.  These results are {\it formally} the same as we found above by introducing a congruence of timelike worldlines with $D$-velocity field $v$, but with the difference that $u$ is determined, given a foliation of the worldvolume,  by the solution of the brane equations. Assuming that $A^2\ge0$, application of the Unruh formula gives
\be\label{UnruhBrane}
T= \frac{|A|}{2\pi}= \frac{1}{2\pi}\,  \sqrt{K_{uu}^2 + a^2}\, ,
\ee
for the brane temperature.  

\subsection{Limitations}

There are several reasons why the formula (\ref{UnruhBrane})  cannot be universally valid. 
To begin with, what we mean by `the brane' depends on how the worldvolume is foliated by spacelike hypersurfaces. This foliation provides a preferred class of observers, those on orbits of  the vector field $u$, but $u$ depends on the choice of foliation because it does not transform as a vector field under general worldvolume coordinate transformations. Specifically, an infinitesimal change of worldvolume coordinates that induces   
$X\to X+\zeta^i\partial_i X$ leads to $u\to u+ \delta u$, where
\be\label{non-vector}
\delta u^i\partial_i  =\left[ \zeta,u\right]^i \partial_i - h^{ab}\partial_b \zeta^t\Delta \, \partial_a\, , 
\ee
where the first term, involving the  commutator of the vector fields $\zeta$ and $u$, is the standard 
infinitesimal transformation of a worldvolume vector field.  Thus, $u$ fails to transform as a vector 
field under space-dependent time reparametrizations, and the brane temperature could depend
on precisely what is meant by `the brane'.  However, temperature is an equilibrium concept so we should not attempt to interpret brane acceleration as temperature unless the worldvolume metric is stationary,  or nearly so. This means that there should be some timelike worldvolume vector field 
$\xi$ that is, at least approximately, Killing. In such cases  there is a `preferred' foliation of the worldvolume: that for which $u= \xi/\sqrt{-\xi^2}$.  In some of the examples that we shall meet, one may choose the foliation such that
\be\label{uxi}
u = \Delta^{-1} \xi\, , 
\ee
where the timelike vector field $\xi$ is either Killing or asymptotically Killing, and the Tolman law (\ref{Tolman}) becomes
\be\label{Tolman2}
T= T_0/\Delta\, . 
\ee

In some of the examples we consider in this paper, the local Unruh temperature, given by  (\ref{UnruhBrane}), takes the form (\ref{Tolman2}), consistent with thermal equilibrium. In other examples
the local Unruh temperature does not take this form and is therefore  not consistent with thermal equilibrium, but in all such cases the Unruh formula turns out to be 
inapplicable. Recall that  the Unruh formula  was derived assuming constant (time-independent) proper acceleration. It is physically reasonable to allow the ($D$-vector) acceleration to change slowly with time but then we need a measure of how slow this change is.  In non-relativistic particle mechanics the time rate of change of acceleration is called `jerk'. In the context of a  congruence of timelike worldlines that are orbits of a vector field $v$, as discussed above,  the natural relativistic generalization of jerk would be $j= v^jD_j a^i(v)$, but this is non-zero even when $a^2$ is constant because $v\cdot j = -a^2$. This suggests that \be\label{reljerk}
\varsigma ^k= j^k -a^2 v^k\, , \qquad j^k= v^i D_i a^k
\ee
is a better definition of relativistic jerk, since $a\cdot \varsigma \equiv a\cdot j$ but $v\cdot \varsigma \equiv 0$.  There is a similar problem with $u^i\partial_i A$ as the definition of  `brane jerk',  which is 
resolved by considering instead the $D$-vector valued  worldvolume field
\be\label{Sigmadef}
\Sigma^\mu = u^i\partial_i A^\mu -A^2 U^\mu\, . 
\ee
We will argue in the following section that
\be\label{smallparameter}
\lambda = \sqrt{\Sigma^2}/A^2\, 
\ee
is  the relevant dimensionless parameter, in the sense that (\ref{UnruhBrane}) should be valid  
whenever $\lambda\ll1$. For all the examples we consider, the local Unruh temperature  
is consistent with (approximate) thermal equilibrium whenever this condition is satisfied.

\subsection{Organization}

In the following section we present further aspects, and technical details, of the above ideas. We then consider a series of illustrative examples. We start with an open $p$-brane in a $(2p+1)$-dimensional Minkowski spacetime, rotating freely, and rigidly, in $p$ orthogonal planes. This is a case in which the (constant but non-uniform) brane acceleration  is everywhere tangential to the (static) worldvolume.  The local Unruh temperature is {\it not} of the Tolman form but 
we also find that the parameter $\lambda$ of (\ref{smallparameter}) is everywhere greater than unity, whereas the Unruh formula is valid only if $\lambda\ll1$. 

Most of our other examples  involve a spherical $p$-brane, possibly coupled to a constant uniform electric-type field, expanding or contracting  in a maximally symmetric spacetime. We consider first the case in which the maximally-symmetric spacetime is $(p+2)$-dimensional Minkowski spacetime. In such cases the worldvolme has co-dimenson one in the ambient flat spacetime, and the brane may be viewed as a spherical domain wall.  For a non-zero electric-type field, there is an equilibrium solution of the brane equation of motion (\ref{eofm}) for which  the external force is cancelled by the `internal' force due to tension; this solution has $A\equiv 0$ and its brane temperature is therefore zero.  The equilibrium is unstable because a small uniform perturbation leads either to collapse or to 
a runaway solution in which the brane expands indefinitely. This runaway solution approaches an asymptotic solution with non-zero
uniform $A$, and $\lambda=0$; its worldvolume geometry is de Sitter and the 
brane temperature is the Gibbons-Hawking temperature.

We then turn to spherical $p$-branes  undergoing constant uniform acceleration in either $dS_{p+2}$ or $adS_{p+2}$.  
As $(a)dS_{p+2}$ may be globally embedded in a $(p+3)$-dimensional flat spacetime, we are effectively considering  motion in a flat spacetime 
but now with a worldvolume that has  co-dimension two rather than co-dimension one. Consider the  special case of a zero-brane 
undergoing constant acceleration $a$ in $dS_2$. The formula (\ref{thirring}) applies, and this can also be found by an application of 
the Unruh formula to the zero-brane's acceleration $A$ in ${\rm Minkowski}_3$, since the two non-zero components of $A$ are $a$ and  $1/R$. 
 From the above discussion of brane acceleration for $p>0$, it should be clear that the same argument  applies to a $p$-brane undergoing 
constant uniform acceleration in $dS_{p+2} \hookrightarrow {\rm Minkowski}_{p+3}$, 
and hence that the formula (\ref{thirring}) still applies. This makes sense because the constant uniform acceleration implies that the worldvolume geometry is de Sitter. Note, however, that $a$ {\it is now to be interpreted as the brane's acceleration in the ambient (a)dS spacetime, rather than the worldvolume acceleration of some test particle on the brane}.  Similarly, the formula (\ref{acceladS}),  and hence (\ref{DL}) when $a\ge 1/R$, applies to a zero-brane  in 
$adS_2$  but generalizes to a $p$-brane in $adS_{p+2}$. This again makes sense because, as we show,  constant uniform acceleration $a$ implies that the worldvolume geometry is $adS_{p+1}$  if $a<1/R$, ${\rm Minkowski}_{p+1}$ if $=1/R$, and $dS_{p+1}$ if $a>1/R$. This is a purely kinematical statement but all these cases occur as solutions to the equations of motion of a $p$-brane coupled to a constant uniform electric-type field; one finds that $a<1/R$ for a `sub-critical' electric field, $a=1/R$ for critical electric field, and $a>1/R$ for super-critical electric field. 

Our final example is the BTZ black hole \cite{Banados:1992wn,Banados:1992gq}. The global embedding of the BTZ metric in a flat $(2+2)$-dimensional spacetime was given in \cite{Deser:1998xb}, 
where it was used to discuss thermal properties, although the local temperature used there is the non-equilibrium temperature of ZAMOs (zero angular momentum observers), which fails to agree with
the Unruh temperature except near the horizon. Our results, obtained by viewing the BTZ black hole as a brane, yield similar results because our definition of brane acceleration effectively picks out the ZAMOs as `preferred' observers. We show that $\lambda$ is small for such observers only near the horizon. 
However, we further show that observers who are rotating with the same angular velocity as the black hole horizon have $\lambda=0$ everywhere, and their Unruh temperature precisely matches the equilibrium black hole temperature found by applying the Tolman law to the timelike Killing vector field
that becomes null on the horizon. The `jerk-free' observers are those on orbits of this Killing vector field. 

We leave to a final section some further discussion of  the implications our results.

\section{Brane Kinematics}

Here we give further details related to some of the ideas discuused above. 
We begin with a derivation of (\ref{BraneA}) from the Hamiltonian formulation, and 
consider in more detail the issue of coordinate-dependence of the brane acceleration.
We then recall some formulae of extrinsic geometry, and investigate the circumstances
under which the brane acceleration can be expressed in terms of  intrinsic worldvolume 
geometry. Finally we elaborate on the concept of  brane jerk and explain its relevance
to the validity of the Unruh formula. 

\subsection{Hamiltonian formalism}

For simplicity, we suppose that the $p$-brane is a closed $p$-surface, and that ${\cal C}=0$; in which case the Hamiltonian form of the action is
\be\label{hamact}
S = \int dt \oint d^p\sigma \left\{\dot X\cdot P - \frac{1}{2} \ell \left( P^2   + \mu^2 \det h \right) - s^a P\cdot \partial_a X\right\} \, , 
\ee
where $(\ell, s^a)$ are the `lapse' and `shift' Lagrange multipliers for the Hamiltonian and worldspace diffeomorphism constraints; note that the latter implies that the $D$-momentum  $P$ is everywhere orthogonal to the brane. To verify the equivalence to the Dirac form of the action, first eliminate $P$ by its equation of motion
\be\label{convective}
P= \ell^{-1}\left(\dot X - s^a\partial_a X\right) = \mu \sqrt{\det h} \, u^i\partial_i X\, , 
\ee
where the second equality follows from the definition 
\be
u^i\partial_i \equiv \frac{\ell^{-1}}{\mu\sqrt{\det h}}\left(\partial_t - s^a\partial_a\right)\, . 
\ee
If we substitute for $P$ in the action and vary the resulting action with respect to $u$, we
find that
\be\label{lagmulteq}
u= \Delta^{-1}\left(\partial_t -s^a\partial_a\right)\, , \qquad s^a=h^{ab}g_{tb}\, , 
\ee
where we recall from (\ref{uform}) that 
\be
\Delta \equiv \sqrt{-g_{tt}+ g_{ta}h^{ab}g_{tb}} ={1\over  \sqrt{-g^{tt}} }\, . 
\ee
One may now verify that the dual worldvolume one-form is given by (\ref{uform}). 
When this result is used to eliminate $u$ from the action,  the action (\ref{Dirac}) (with ${\cal C}=0$) is recovered upon use of the identity 
\be\label{detid}
\sqrt{-\det g} = \Delta \sqrt{\det h}\, . 
\ee
As noted in the introduction,  $u^2=-1$, from which follows the interpretation of $u$ as a worldvolume velocity field. Also, if  (\ref{lagmulteq}) is used in (\ref{convective}) then we recover  the formula (\ref{PU}) for brane momentum density, with the brane velocity $U$ given by (\ref{UUU}).

Returning to the action (\ref{hamact}), we observe that the $X$ equation of motion is
\be\label{Xeq}
\dot P - \partial_a \left(s^aP\right) = \mu^2 \partial_a\left(\ell \det h h^{ab}\partial_b X\right) \ .
\ee
Using (\ref{lagmulteq}) and (\ref{detid}), we may rewrite this as
\be\label{Newton's2}
\dot P - \partial_a \left(s^aP\right) = \sqrt{-\det g}\, F_{int}\, , 
\ee
where the `internal' force density, due to the brane tension, is
\be
\sqrt{-\det g}\ F_{int} \equiv  \mu\, \partial_a\left(\sqrt{-\det g}\, h^{ab}\partial_b X\right)\, .  
\ee
Now we use (\ref{convective}), and then  (\ref{lagmulteq}), to deduce that
\be
\dot P - \partial_a \left(s^aP\right) =  \mu\,  \partial_i\left(\sqrt{-\det g}\, u^i U\right)\, .
\ee
Equivalently,
\be
\dot P - \partial_a \left(s^aP\right) = \mu \sqrt{-\det g}\, A + 
\mu\, \partial_i\left(\sqrt{-\det g}\, u^i\right) U\, , 
\ee
where $A=u^i\partial_i U$ is the brane acceleration as defined in the introduction. We now see that 
(\ref{Newton's2}) is equivalent to 
\be
F_{int} = \mu  A +  \mu \left(D_iu^i\right) U\, . 
\label{neww}
\ee
In the presence of an external force, this equation still holds but with
\be
F_{int} \to F_{int} + F_{ext} \equiv F \, . 
\ee
Equation (\ref{neww}) is essentially Newton's second law. For $p=0$ it reduces 
to the standard form of this law for a relativistic particle of rest-mass 
$\mu$. For $p>0$ there is an extra term proportional to the covariant 
divergence of $u$, as a consequence of which we deduce for a closed brane 
that
\be\label{Mdot}
\oint \!d^p\sigma \, \Delta\, P\cdot F_{int} = -\mu\dot M\, , \qquad M\equiv \mu \oint\!d^p\sigma \sqrt{\det h}\ .
\ee
It is instructive to compare this with the example of a relativistic particle with variable rest-mass $m(t)$ and velocity $v(t)$, and hence momentum $p=mv$. Given that $F= dp/d\tau$, where $\tau $ is proper time, one has
$p\cdot F = -m\ dm/d\tau$, in close analogy with (\ref{Mdot}). The reason should be clear: a non-vanishing divergence of the velocity field implies a non-conservation  of  {\it local} rest-mass. Globally, the brane behaves like a particle at its centre of mass, and the rest-mass of this `particle' is conserved because it  includes the kinetic energy of the brane.  

\subsection{Extrinsic vs Intrinsic}

We first recall some basic formulae of extrinsic geometry.  
We suppose that a $(p+1)$-dimensional spacetime, which we call the `worldvolume', is globally embedded in a flat $D$-dimensional spacetime, which may have more than one time but is  such that the induced worldvolume metric has Lorentzian signature.  We may write the flat $D$-metric as
\be
ds^2_D = \eta_{\mu\nu}dX^\mu dX^\nu = -(dX^0)^2 + (dX^1)^2 + \cdots
\ee
where the dots indicate additional dimensions, one of which may be an additional time dimension. 
Let $\{\xi^i; i=0,1,\dots,p\}$ be local worldvolume coordinates. The induced worldvolume metric in these coordinates is 
\be
 g_{ij} = \eta_{\mu\nu} \partial_i X^\mu \partial_j X^\nu\, . 
\ee
The extrinsic curvature of the worldvolume is the $D$-vector valued symmetric worldvolume  tensor
\be
K_{ij}^\mu = \partial_i \partial_j X^\mu - \Gamma_{ij}{}^k \partial_k X^\mu\, , 
\ee
where $\Gamma_{ij}{}^k$ is the Levi-Civita connexion for the induced worldvolume metric. This extrinsic curvature tensor satisfies the `Brane Bianchi' identity  (\ref{BBI}), which  implies that the only non-zero components of $K$ are those orthogonal to the worldvolume (but this could include a tangential component if  the flat $D$-dimensional spacetime has more than one time dimension).

It is now straightforward to derive the  formula (\ref{altA}) for $A$, starting from 
the formula
\be
A^\mu=u^i \partial_i U^\mu = u^i\partial_i \left(u^j\partial_jX^\mu\right)\, . 
\ee
In the special case that $a=0$, we find that
\be\label{AKuu}
A^\mu= K_{uu}^\mu \equiv u^iu^jK_{ij}^\mu\, . 
\ee
Thus, in such cases, $A$ is defined in terms of the extrinsic worldvolume geometry. We now aim to investigate the conditions under which  it can be re-expressed entirely in terms of the {\it intrinsic} worldvolume geometry.  We begin by observing that   (\ref{GC1}) implies the following formula for the Ricci tensor:
\be\label{GC2}
R_{ij} = K_{ij}\cdot  \left(g^{kl}K_{kl}\right) - g^{kl} K_{ik} \cdot K_{lj}\, .  
\ee
Using the brane equation of motion (\ref{eofm}), and eq. (\ref{inverseg}), we can rewrite this as
\be\label{GCN}
R_{ij} = -K_{ij}\cdot F -  h^{ab} K_{ai}\cdot K_{bj}  +  u^ku^l K_{ki} \cdot K_{lj} \, . 
\ee
By contracting with $u^i u^j$, we find that
\be
R_{uu}\equiv u^iu^jR_{ij} = K_{uu}^2- K_{uu}\cdot F - 
 h^{ab} \left[u^iK_{ia}\right]\cdot \left[u^jK_{jb}\right] \, .   
\label{kio}
\ee
It follows that a {\it necessary} condition for $K_{uu}^\mu$ to be expressible in terms the intrinsic geometry is 
that  
\be
\label{ooi}
h^{ab} \left[u^iK_{ia}\right]\cdot \left[u^jK_{jb}\right]=0 
\ee
This is satisfied, in particular, for\footnote{Remarkably, this is also the condition for $K_{uu}^\mu$ to transform, infinitesimally, as a scalar under reparametrizations of the worldvolume.}
\be\label{uk2}
u^iK_{ia}^\mu=0\, . 
\ee
When the world-volume is co-dimension one, the condition (\ref{uk2}) is equivalent to (\ref{ooi}).
In this case
\be\label{ARF}
\left(2K_{uu} - F\right)^2 = 4R_{uu} + F^2 \, . 
\ee
In principle the  $D$-vector field $(2K_{uu}-F)$ could depend on the extrinsic geometry in such a way that the magnitude is independent of the extrinsic geometry. In the case of a worldvolume of co-dimension one this cannot happen because then $(2K_{uu}-F)$ has only one component. 

\subsection{Brane Jerk} 

Jerk is the time rate of change of acceleration. The $D$-vector valued  worldvolume field 
$u^i\partial_iA$ is the natural definition of `brane jerk'  but  this does not vanish when 
$A^2$ is constant. We therefore define
\be
\Sigma = u^i\partial_i A - A^2 U\, . 
\ee 
As a consequence of the identity $U\cdot A\equiv 0$, we have the further identity
\be
U\cdot \Sigma \equiv 0\, . 
\ee
We can decompose $\Sigma$ into a component orthogonal to the worldvolume and 
a component tangent to it:
\be
\Sigma^\mu=\Sigma_\perp^\mu +\varsigma^i\partial_i X^\mu\ ,
\ee
where $\varsigma^i$ is the world-volume jerk as defined in (\ref{reljerk}), and
\be 
\Sigma_\perp^\mu = u^iu^ju^k \left[ D_iK_{jk}^\mu -\partial_k X^\mu  \left( A\cdot K_{ij}\right )\right]+ 3a^iu^jK_{ij}^\mu\, . 
\ee

An interesting example is the case of maximally symmetric spaces of radius $R$, for which 
\be
K_{ij}^\mu={1\over R} g_{ij} n^\mu \ ,\qquad  n^2=\pm 1\ ,
\ee
where the  plus (minus)  sign corresponds to (anti) de Sitter. In this case 
\be
\partial_i n^\mu = -{n^2\over R}\partial _i X^\mu\ .
\ee
Using these expressions, one finds that the orthogonal component of $\Sigma $ is identically zero 
and hence that $\Sigma^\mu = \varsigma^i\partial_i X^\mu $. Therefore $\varsigma = 0$ implies $\Sigma =0$ for maximally symmetric world-volumes.

For the purposes of the Unruh formula, the time rate of change of the acceleration and hence of the 
temperature, will be ``small'' if  $|\Sigma|=\sqrt{\Sigma^2}$  is small. 
Given $A^2>0$, we expect $|A|=\sqrt{A^2}$ to be the typical frequency associated with the Unruh 
temperature (in fundamental units), and this should be much greater that $\dot T/T$. 
This amounts to requiring 
\be\label{val}
\lambda \equiv |\Sigma|/A^2 \ll 1\, . 
\ee

\section{Open rotating branes}

Consider a $(2p+1)$-dimensional Minkowski space
\be
ds^2 = -dT^2 + \sum_{\alpha=1}^p\left[ dX_\alpha^2 + dY_\alpha^2\right]
\ee
Now set
\be
T=t\, , \qquad
X_\alpha = \sigma_\alpha\cos\left(\phi_\alpha - \omega_\alpha t\right)\, , \qquad
Y_\alpha = \sigma_\alpha\sin\left(\phi_\alpha - \omega_\alpha t\right)\, . 
\ee
The Minkowski metric in the new coordinates $(t;\sigma_1,\phi_1; \dots; \sigma_p,\phi_p)$ is
\be
ds^2= - \left(1-\sum_{\alpha=1}^p \omega_\alpha^2 \sigma_\alpha^2\right) dt^2 + 
\sum_{\alpha=1}^p \left[\sigma_\alpha^2  d\phi_\alpha^2
- 2\omega_\alpha \sigma_\alpha^2  d\phi_\alpha dt  + d\sigma_\alpha ^2\right]\, .
\ee
We now have the Minkowski metric in coordinates adapted to a frame rotating in $p$ orthogonal planes with angular velocity vector $(\omega_1,\dots,\omega_p)$. The vector field 
\be
k= \partial_t
\ee
is a Killing vector field, but it is not the same Killing vector field as $\partial_T$:
\be
\partial_T = \partial_t + \sum_{\alpha=1}^p \omega_\alpha \frac{\partial}{\partial \phi_\alpha}\, . 
\ee
In contrast to $\partial_T$, the vector field $\partial_t$ has a horizon, on the ellipsoid 
\be\label{ellipsoid}
\sum_{\alpha=1}^p \omega_\alpha^2\sigma_\alpha^2=1\, . 
\ee
The metric is singular on this ellipsoid, but this is, of course,  just a coordinate singularity. 

Now consider an open $p$-brane in this spacetime, and identify $(t;\sigma_1,\dots \sigma_p)$ with the worldvolume coordinates, and suppose further that it is rigidly rotating, so that all $p$ angles 
$\phi_\alpha$ are constant. It is straightforward to verify that this is a solution of the equation of motion, making use of the fact that  the induced worldvolume metric is
\be
ds^2_{ind} =  - \Delta^2 dt^2  + \sum_{\alpha=1}^p d\sigma_\alpha ^2\, , \qquad
\Delta = \sqrt{1-\sum_{\alpha=1}^p \omega_\alpha^2 \sigma_\alpha^2}\, . 
\ee
This metric is static, with Killing vector field $\partial_t$, but has a curvature singularity 
on the $(p-1)$-ellipsoid (\ref{ellipsoid}), which we identify as the boundary of the $p$-brane. All points 
on this boundary move at the speed of light.  From the induced metric we see that
\be
u= \frac{1}{\sqrt{1-\sum \omega_\alpha^2 \sigma_\alpha^2}}\, \partial_t
\ee
Further direct computation shows that
\be
a= - \Delta^{-2} \sum_{\alpha=1}^p \omega_\alpha^2\sigma_\alpha 
\frac{\partial}{\partial\sigma_\alpha}
\ee
and also that 
\be \varsigma = \Delta^{-5}\left( \sum_{\alpha=1}^p 
\omega_\alpha^4\sigma_\alpha^2\right)\partial_t - a^2 u = 0\, . 
\ee 
The last 
equality, which is expected from the time-independepence of $a^2$, shows 
that the brane jerk $\Sigma$, which we compute below, is orthogonal to the 
worldvolume.

A similar computation of the full brane velocity and acceleration yields the result that
$U=u$ and $A=a$, as vector fields, so this is a case in which the brane acceleration is purely 
tangential; equivalently, $K_{uu}=0$, as a computation confirms. An application of the Unruh formula would give 
\be
2\pi \, T= \Delta^{-2} \sqrt{\sum_{\alpha}\omega_\alpha^4\sigma_\alpha^2}\, ,
\ee
but this is not of the form (\ref{Tolman}) required for thermal equilibrium. This feature could have been 
anticipated from the fact that the null hypersuface swept out by the brane's  boundary  is a curvature singularity of the worldvolume metric. Physically, we should expect the brane to radiate as a consequence of its acceleration, and there is no guarantee that thermal  equilibrium can be achieved by immersion in a heat bath. In view of this, we should ask whether application of the Unruh formula is justified in this case. 

A direct computation of the brane jerk, shows that
\be
\Sigma = \Delta^{-3}\sum_{\alpha=1}^p \omega_\alpha^3\frac{\partial}{\partial\phi_\alpha}
- \Delta^{-5} \left(\sum_{\alpha=1}^p\omega_\alpha^4\sigma_\alpha^2\right) \partial_t\, , 
\ee
which yields
\be\label{testvalidity}
\lambda^2 = 
\frac{\sum_\alpha \omega_\alpha^6\sigma_\alpha^2}{\left(\sum_\alpha \omega_\alpha^4\sigma_\alpha^2\right)^2} + \left[ 1 - \frac{\left(\sum_\beta\omega_\beta^2\sigma_\beta^2\right)\left(\sum_\alpha \omega_\alpha^6\sigma_\alpha^2\right)}{\left(\sum_\gamma \omega_\gamma^4\sigma_\gamma^2\right)^2}\right] 
\ee
where $\lambda$ is the parameter of  (\ref{val}). This can be shown to be equivalent to
\be
\lambda^2 \left(1-\Delta^2\right) =
1 + \frac{\Delta^2}{\left(\sum_\gamma\omega_\gamma^4\sigma_\gamma^2\right)^2}
\sum_{\alpha\ne\beta} \left[\left(\omega_\alpha^2-\omega_\beta^2\right)\omega_\alpha\omega_\beta\sigma_\alpha\sigma_\beta\right]^2\, . 
\ee
It follows that
\be
\lambda^2 \ge  \left(1-\Delta^2\right)^{-1} \ge 1\, , 
\ee
where the second inequality follows from the fact that $\Delta^2<1$. This violates  (\ref{val}), so there is no region of the brane in which we can use the Unruh formula!


\section{Spherical branes in flat space}

A simple class of solutions of the $p$-brane equations of motion (\ref{eofm}) is obtained 
by considering the motion of a spherical $p$-brane in a $(p+2)$-dimensional Minkowski spacetime. 
In this case we can think of the brane as an effective description of a spherical domain wall, with a worldvolume that is a hypersurface in the Minkowski  spacetime.  In the absence of an external force,
all solutions of (\ref{eofm}) describe  minimal Lorentzian surfaces in which a spherical brane oscillates between expansion and contraction due to the `internal'  force provided by its tension. 
The inclusion of an external, constant and uniform, force allows for a static, but unstable, spherical  brane with a definite, equilibrium radius of curvature. A radial perturbation causes such a brane to either collapse or to expand indefinitely, in which case it approaches an asymptotic solution for which the worldvolume geometry is de Sitter. 

\subsection{Kinematics}

As these examples are all to do with the motion of a spherical $p$-brane, it is convenient to 
choose spherical polar coordinates for the flat ambient space, such that the Minkowski spacetime metric 
takes the form
\be
ds^2 = -dt^2 + dr^2 + r^2 d\Omega_p^2\, , 
\ee
where $d\Omega_p^2$ is the unit $SO(p+1)$-invariant metric on the $p$-sphere. As the notation suggests, we choose $t$ to coincide with worldvolume time. We may also choose the brane coordinates to coincide with the angular coordinates of the $p$-sphere.  If the brane has radius $r(t)$ at time time $t$ then  the induced worldvolume metric is
\be\label{wmetric}
ds^2_{ind}=  -\left(1-\dot r^2\right) dt^2 + r^2 d\Omega_p^2\, , 
\ee
where the overdot indicates differentiation with respect to $t$.  
It follows that the worldvolume velocity field $u$ is
\be\label{littleU}
u= \Delta^{-1} \partial_t\, , \qquad \Delta = \sqrt{1-\dot r^2}
\ee 
and a straightforward calculation shows that the worldvolume acceleration is zero. The brane acceleration is therefore orthogonal to the worldvolume in these examples. Note that $\partial_t$ is
not Killing unless $\dot r=0$, so only in this case does  $u$ takes the form (\ref{uxi}).  

Using (\ref{littleU}) one may deduce that $U$ and $A$, as vector fields on Minkowski spacetime, are given by
\be
U= \Delta^{-1} \left(\partial_t + \dot r \partial_r\right)\, , \qquad
A = \ddot r \Delta^{-4} \left(\dot r \partial_t + \partial_r\right)\, .
\ee
It follows that 
\be\label{Asquared}
A^2 = \frac{\ddot r ^2}{\left(1-\dot r^2\right)^3}\, .  
\ee
One can similarly compute the brane jerk:
\be
\Sigma = \Delta^{-3}\left(\dddot r + 3\Delta^{-2}\ddot r^2 \dot r\right)
\left(\dot r \partial_t + \partial_r\right)\, .
\ee
This leads to 
\be\label{vlid}
\lambda
= \frac{1-\dot r^2}{\ddot r^2}\left[ \dddot r \left(1-\dot r\right)^2 + 3\ddot r^2 \dot r\right] =
\frac{\left(1-\dot r^2\right)^2}{2\ddot r} A^{-2}d A^2/dt\, . 
\ee
As expected, $\lambda=0$ when $\dot r\equiv 0$, and in this case the brane temperature is well-defined, but zero. In addition $\lambda=0$ when $A^2$ is constant, and in such cases the 
local Unruh temperature is
\be\label{TU}
T= \frac{\left|\ddot r\right|}{2\pi\left(1-\dot r^2\right)}\, . 
\ee
This would appear to be inconsistent with  (\ref{Tolman2}), but here we should recall that  the vector field $\xi$ defined by $u= \Delta^{-1}\xi$ is not Killing  unless $\dot r=0$, so that (\ref{Tolman2}) cannot be applied in the coordinates that we are using.  It could be that  there are other coordinates such that 
$u=\Delta^{-1}\xi$ for Killing vector field $\xi$,  so that  (\ref{Tolman2})  is applicable. As we shall see, 
this is the case, and (\ref{TU}) is consistent with (\ref{Tolman2}) in these new coordinates.

\subsection{Dynamics}

For our discussion of the dynamics, we will wish to allow for a  uniform electric-type field in the Minkowski background; i.e. 
\be
F_{p+2} = E r^p dt \wedge dr\wedge dV_p
\ee
where $E$ is a constant and $dV_p$ is the volume element on the $p$-sphere. We may choose a gauge in which the $(p+1)$-form potential is
\be
C_{p+1} = -\frac{E}{p+1} r^{p+1} dt\wedge dV_p\, , 
\ee
in which case the worldvolume Hodge dual of the $(p+1)$-form induced by $C_{p+1}$ is
\be
{\cal C}= -\frac{E}{p+1} r^{p+1}\, . 
\ee

Putting things together, we see that the action (\ref{Dirac}) becomes
\be
S[r] = \mu V_p \int  dt \left\{ -r^p\sqrt{1-\dot r^2} + \frac{E}{(p+1)} r^{p+1}\right\}\, , 
\ee
where $V_p$ is the volume of the unit $p$-sphere. The Lagrangian has no explicit dependence on $t$ so there is a corresponding conserved quantity ${\cal E}$, and the first integral of the equations of 
motion  is 
\be\label{firstint}
\dot r^2 = 1- \frac{(p+1)^2 r^{2p}}{\left({\cal E} +Er^{p+1}\right)^2}\ .
\ee
Using this equation to eliminate $\dot r$ from (\ref{Asquared}), we find that 
\be
A^2= {1\over (p+1)^2}\left(E-{ p{\cal E}\over r^{p+1}} \right)^2\, . 
\label{xer}
\ee
We may also use (\ref{firstint}) to deduce from (\ref{vlid}) that
\be\label{simplevalid}
\lambda =
\frac{p\left(p+1\right)^3r^{2p}\,  {\cal E} \dot r}{\left({\cal E} + Er^{p+1}\right)
\left(Er^{p+1} -p{\cal E}\right)^2}\, . 
\ee
Notice that this vanishes in precisely two cases: (i) $\dot r=0$ and (ii) ${\cal E}=0$.

\subsection{Zero external force}

In the absence of an electric field, i.e. $E=0$,  we can rewrite (\ref{firstint}) as
\be
\dot r^2 = 1 - \left(r/r_{\rm max}\right)^{2p}\, , 
\label{juno}
\ee
where  the integration constant  $r_{\rm max}$  is the maximum value of $r$.   For $p=1$ the solution is $r=r_{\rm max} \sin t $, and quite generally we may choose $t$ such that $r=0$ at $t=0$.  Even without having to integrate, we can substitute for $\dot r$ in (\ref{wmetric}) to see that the worldvolume metric is 
\be\label{worldvol}
ds^2_{p+1}= - \left(r(t)/r_{\rm max}\right)^{2p} dt^2+r^2(t) d\Omega_p^2 \, .
\ee
In this case we may compute the brane acceleration directly from the `intrinsic' formula (\ref{ARF}) for $F=0$,  which  yields 
\be 
A^2 = R_{uu}=  -R_{tt}/g_{tt} = {|\ddot r|^2\over (1-\dot r^2)^{3}} = \left({p \ r_{\rm max}^p\over r^{p+1}}\right)^2\, , 
\label{asse}
\ee
in agreement with  (\ref{xer}), for $E=0$.  From (\ref{simplevalid}) we find that
\be
\lambda = \frac{p+1}{p}\left(\frac{r}{r_{max}}\right)^{2p}
\sqrt{1- \left(\frac{r}{r_{max}}\right)^{2p}}\, . 
\ee
We have $\lambda\ll1$ as $r\to r_{max}$, so the Unruh formula should apply, and we deduce that the brane has a temperature $T\approx p/(2\pi r_{\rm max})$. We also have $\lambda\ll1$ as $r\to0$, 
but in this case the temperature is rapidly increasing without bound and we expect  other physics to intervene.

\subsection{Non-zero external force}
\label{subsection:nonzero}

For positive $E$, there is an unstable static solution of (\ref{firstint}) with
\be
r= p/E\equiv r_0 \, , \qquad {\cal E}= r_0^p\, . 
\ee
The induced metric is that of an Einstein Static Universe:
\be
ds^2_{ind} = -dt^2 + r_0^2 d\Omega_p^2\, . 
\ee
The  brane acceleration according to (\ref{xer}) is $A=0$, in agreement with (\ref{ARF}) since  
 $R_{uu}=0$ in this case. Since $\dot r=0$, we get $\lambda=0$ from  (\ref{simplevalid}). The
 Unruh formula applies in this case but the brane has zero temperature. 
 
 \subsubsection{The de Sitter solution}
 
 All solutions of (\ref{firstint}) for which $r\to\infty$ are asymptotic to the solution 
\be
r= \sqrt{R^2+ t^2} \, , \qquad {\cal E}=0\, \qquad (R = (p+1)/E)\, , 
\ee
for which the induced metric is
\be
ds^2_{ind} = - \frac{R^2}{R^2+t^2} dt^2 + \left(R^2+t^2\right)d\Omega_p^2\, . 
\ee
For $p=1$, ths solution was found in \cite{Dowker:1995sg}; apart from the extension to $p>1$, the main new observation here is that the worldvolume geometry of this asymptotic solution is de Sitter. To see this, we substitute the solution  into (\ref{wmetric})  and then introduce a new time coordinate 
$\tau$ such that
\be
t = R \sinh \tau\, . 
\ee
The resulting worldvolume metric is 
\be\label{dsinduced}
ds^2_{ind} =R^2 \left[ -d\tau^2 + \left(\cosh\tau\right)^2 d\Omega_p^2\right]\, , 
\ee
which is a standard parametrization of the metric of a de Sitter spacetime with radius of 
curvature $R$.  Note that 
\be
u^i \partial_i = R^{-1} \partial_\tau
\ee
so the brane velocity and acceleration are effectively defined as first and second derivatives with respect to the time 
parameter $\tau$.  From (\ref{xer}) we see that, for this solution
\be
A^2 = 1/R^2\, , 
\label{lop}
\ee
so the brane acceleration is constant, with magnitude $1/R$. Since $\lambda=0$ we may apply the
Unruh formula to deduce that the brane temperature is $T= 1/(2\pi R)$, which is  
the Gibbons-Hawking temperature of de Sitter space. 

There is another choice of coordinates for which the de Sitter  spacetime is static:
\be
ds^2_{p+1} = -\left(1-\frac{\tilde r^2}{R^2}\right) d{\tilde t}^2 + \frac{R^2 d\tilde r^2}{R^2- \tilde r^2}
+  {\tilde r}^2 d\Omega_{p-1}^2\, . 
\ee
The coordinate singularity at $\tilde r=R$ is the event horizon of an observer at $\tilde r=0$. The Killing vector field $k= \partial_{\tilde t}$ is such that $k^2=-1$ at this observer, who therefore plays a role analogous to that of an observer at infinity in a black hole spacetime. The Euclidean metric with 
$\tilde t = iR\theta$ is non-singular at $\tilde r=R$ if  $\theta$ is identified with $\theta+ 2\pi$, which means that the metric is periodic in imaginary time with period $2\pi R$, which is interpreted as the inverse of the (Gibbons-Hawking) temperature of the observer at the origin \cite{Gibbons:1977mu}:
\be
T_{GH} = \frac{1}{2\pi R}\, . 
\ee
In these static coordinates for the de Sitter metric, the worldvolume velocity field is again $u= \Delta^{-1}\partial_{\tilde t}$, but now $\Delta= \sqrt{1- \tilde r^2/R^2}$ and $\partial_{\tilde t}$ is a Killing vector field.

In static coordinates, de Sitter spacetime does not have an obvious interpretation as a brane, but
it is still embedded in Minkowski spacetime of one higher dimension. The embedding is
\be
X_0 \pm X_{p+1} = \pm \sqrt{R^2-\tilde r^2}\, e^{\pm \tilde t/R} , \qquad \left(X_1, \dots , X_p\right) 
= \left(x_1,\dots x_p\right)
\ee
where $x_1^2 + \dots + x_p^2 = \tilde r^2$. Defining $U$, $A$ and $\Sigma$ as before, one finds that
\be
A^2 = \frac{1}{R^2 -\tilde r^2} 
\label{lopp}
\ee
One also finds that $\Sigma\equiv 0$, so we may apply the Unruh formula to find that
\be
T= \frac{1}{2\pi}\left[\frac{1}{R^2} + a^2(\tilde r)\right] \, , \qquad  a(\tilde r) = \frac{\tilde r}{R\sqrt{R^2-\tilde r^2}}
\ee
where $a(\tilde r)$ is the acceleration within the de Sitter space of an observer at fixed $\tilde r$. 

This example illustrates that a change of coordinates, leading to a change in the world-volume velocity, can change the brane acceleration
$A$ (cf. eqs. (\ref{lop}), (\ref{lopp})). In general, one should expect $\Sigma $ to change too. However, for de Sitter world-volume, we have shown previously that $\Sigma $ is determined
by $\varsigma $ which is zero whenever $a$ is constant, as it is in the above examples.

\subsubsection{Higher-dimensional Rindler spacetime}

The same de Sitter solution may be found by another method. As we shall use this method  extensively in the next section, we illustrate it here.  We write the Minkowski metric, or rather part of it, as a foliation by dS hypersurfaces. This leads to a natural generalization of the $(1+1)$-dimensional Rindler metric. 
Specifcally, we define new coordinates $(\tau,\rho)$  by 
\be
t= \rho\sinh\tau \, , \qquad r= \rho\cosh\tau\, , 
\ee
to get the Minkowski spacetime metric in the form
\be\label{dSsliceMink}
ds^2_{p+2} = d\rho^2 + \rho^2 \left[-d\tau^2 + \cosh^2\tau \, d\Omega_p^2\right]\, .  
\ee
In these coordinates, 
\be
F =\rho^{p+1}(\cosh\tau)^p\, d\rho \wedge d\tau \wedge d\Omega_p\, ,  
\ee
and the effective Lagrangian is 
\be
L_{\rm eff} = -\rho^p (\cosh\tau)^p\sqrt{\rho^2 - (\partial_\tau \rho)^2} 
+ R^{-1}\frac{(p+1)}{p+2} \rho^{p+2}(\cosh\tau)^p\, . 
\ee
This is time-dependent, which complicates the study of general solutions, but if we focus on solutions with $\partial_\tau \rho=0$ then we need only consider the effective potential
\be
V_{\rm eff} \equiv - L_{\rm eff}|_{\dot\rho=0} = (\cosh\tau)^p \left[\rho^{p+1} - 
\frac{p+1}{R(p+2)} \rho^{p+2}\right]\, ,
\label{enne}
\ee
which  is minimized when $\rho=R$.

\subsubsection{General solution}

Let us now consider the general solution, with non-zero  ${\cal E}$, that approaches the dS solution at late times. For such solutions, $r$  increases indefinitely, so that
\be
\dot r \sim 1 - \frac{R^2}{2r^2} + \dots
\ee
where $R=(p+1)/E$ is the radius of the eventual dS spacetime. We then have
\be
|A| \sim R^{-1}\left[1 - \frac{p{\cal E}}{2Er^{p+1}} + \dots\right]\, , \qquad
\lambda \sim \frac{pR^3 {\cal E}}{r^{p+3}} + \dots
\ee
Since $\lambda$ is small and the worldvolume geometry is approximately de Sitter, there is a local temperature given approximately by
\be
T = T_{GH} \left[1-  \frac{p{\cal E}}{2Er^{p+1}} + \dots\right]\, , \qquad T_{GH}= \frac{1}{2\pi R}\
\ee
If we consider a spherical $p$-brane that starts close to the static solution with $r=p/E$ and then expands to approach, asymptotically, the dS solution, we start with zero temperature and end with 
the non-zero Gibbons-Hawking temperature $T_{GH}$. In the intermediate phase it may not make sense to attribute a temperature to the brane but eventually we may consider the brane to have a temperature 
slightly less than $T_{GH}$ which is approached asymptotically.  


\section{Spherical branes in (a)dS}

We now consider a spherical $p$-brane in a maximally  symmetric  $(p+2)$-dimensional spacetime that is not Minkowski; in other words in either $dS_{p+2}$ or $adS_{p+2}$. As these may be globally embedded in a flat $(p+3)$-dimensional spacetime, we could analyse the problem from this perspective, in which case we would be dealing with a spherical brane in a flat spacetime but such that the worldvolume has co-dimension two rather than co-dimension one (as was the case in the previous section). 

We will focus  on those cases for which the worldvolume is itself a maximally symmetric spacetime. Such solutions are possible  in the presence of a constant uniform force produced by a non-zero constant uniform electric-type field, which we henceforth assume.  For  $p=0$ this means that we are considering a particle undergoing constant acceleration $a$ in $(a)dS_2$, and we know from the earlier work summarized in the introduction that   $A^2=a^2 \pm R^{-2}$, where $R$ is the (a)dS radius.  Here we use our concept of brane acceleration to show how this formula also applies to $p$-branes undergoing constant uniform acceleration $a$ in a $(p+2)$-dimension (a)dS space, except that now we must  specify that we consider  the solution for which the worldvolume geometry is maximally symmetric because other solutions are possible.  

Because of this focus on maxmally symmetric worldvolumes, the simplest way to find solutions is to 
consider the possible foliations of (a)dS by a family of (a)dS hypersurfaces, parametrized by some 
radial coordinate, and to seek solutions for which this radial coordinate is constant on the brane's worldvolume. This reduces the problem to looking for stationary points of an effective potential, as 
illustrated in the previous section for the `dS brane' expanding in Minkowski spacetime.

\subsection{ Accelerating brane in  de Sitter space}
\label{subsec:dSndS}

We begin by considering a $p$-brane undergoing uniform acceleration in a $(p+2)$-dimensional dS spacetime of radius $R$. As observed above, we could embed the dS spacetime in an `auxilary'  $(p+3)$-dimensional Minkowski spacetime, and thereby convert the problem into one of  a brane moving in a flat spacetime,  but it is as simple to consider the curved `physical' dS spacetime directly.  A global foliation of dS by leaves that are maximally symmetric is possible only if the leaves of the foliation are also de Sitter spaces, so we take the {\it spacetime} de Sitter metric in the form
\be\label{dSdS}
ds^2_{p+2} =R^2\left[ dy^2+ \cos^2 ({y}) \bar g^{dS}_{ij} d\sigma^i d\sigma^j \right]\, , 
\ee
where $\bar g^{dS}_{ij}$ is the metric on a unit radius de Sitter space of dimension $(p+1)$, with coordinates $\sigma^i$ ($i=0,1,\dots,p$). As the notation suggests, we will identify these coordinates with the worldvolume coordinates  of the $p$-brane. The induced worldvolume metric is then
\be
ds^2_{ind} = R^2 \cos^2y \left[\bar g^{dS}_{ij} + \partial_i y \partial_j y\right]\, d\sigma^id\sigma^j\, . 
\ee
We assume a constant uniform electric-type field of strength $E$, so that
\be
F_{p+2} = ER^{p+2} \cos^{p+1}y \, dy\wedge dV_{p+1}(dS)
\ee
where $dV_{p+1}(dS)$ is the volume $(p+1)$-form of the unit radius dS space. We may choose a gauge  for which the electric $(p+1)$-form field is
\be
C_{p+1} = ER^{p+2} f(y) dV_{p+1}(dS)\, , \qquad f'(y) = \cos^{p+1}y\, . 
\ee
Using these results in (\ref{Dirac}) we find the effective action
\be
S[y] = -\mu R^{p+1}\int dV_{p+1}(dS)\left\{ \cos^p y\sqrt{\cos^2y + 
\bar g_{dS}^{ij}\partial_i y\partial_j y} -ER f(y)\right\}\, . 
\ee
Solutions of the equations of motion with constant $y$ are easily found by extremizing the effective potential 
\be
V_{eff}(y) = \mu R^{p+1} \left[ \cos^{p+1}y -ER\, f(y)\right]\, . 
\ee
There is a maximum of this potential when
\be
\tan y= -\frac{ER}{p+1}
\ee
and in this case the induced worldvolume metric is dS with radius $R_{ind}$ such that
\be
\frac{1}{R_{ind}^2}=  \frac{1}{R^2} + \left(\frac{E}{p+1}\right)^2 
\ee
As the worldvolume is a dS space, its acceleration $A$ in the higher-dimensional flat space is such that
$A^2 = R_{ind}^{-2}$. Applying the Unruh formula, we see that  the brane temperature is given  
\be
2\pi\, T= \sqrt{a^2 + R^{-2}} \, , \qquad a = E/(p+1)
\ee
which is  (\ref{thirring}) but with $a$ interpreted as the acceleration of the $p$-brane in the $(p+2)$-dimensional dS spacetime.  The latter has an acceleration $1/R$ in the $(p+2)$-dimensional  Minkowski spacetime, which provides another way of understanding our result for $A^2$. 

Observe that the brane temperature equals that of the dS spacetime in which it is embedded only 
if $E=0$. In this case the brane is in mechanical equilibrium and thermal equilibrium with its surroundings. A non-zero electric field causes the brane to accelerate in the surrounding dS
spacetime, so in this sense it is no longer in mechanical equilibrium. As we have seen, this alo increases the brane temperature so the brane is also out of thermal equilibrium with the surrounding dS spacetime.

\subsection{ Accelerating branes in anti de Sitter space}

We now  consider  a $p$-brane undergoing uniform acceleration in a $(p+2)$-dimensional adS spacetime of radius $R$. As in the dS case, there exist foliations of adS in which the leaves are maximally symmetric, but  this worldvolume geometry could now be dS, Minkowski, or adS.

\subsubsection{adS worldvolume}

Let us first consider the adS foliations because this case is analogous to the foliation of dS by dS spaces.  We write the {\it spacetime} adS metric in the form
\be\label{dSdS2}
ds^2_{p+2} =R^2\left[ dy^2+ \cosh^2 ({y}) \bar g^{adS}_{ij} d\sigma^i d\sigma^j  \right]\, , 
\ee
where $\bar g^{adS}_{ij}$ is the metric on a unit radius anti-de Sitter space of dimension $(p+1)$, with coordinates $\sigma^i$ ($i=0,1,\dots,p$). As the notation suggests, we will identify these coordinates with the worldvolume coordinates  of the $p$-brane. The induced worldvolume metric is then
\be
ds^2_{ind} = R^2 \left[ \cosh^2y \bar g^{adS}_{ij} + \partial_i y \partial_j y\right]\, d\sigma^id\sigma^j\, . 
\ee
We assume a constant uniform electric-type field of strength $E$, so that
\be
F_{p+2} = ER^{p+2} \cosh^{p+1}y \, dy\wedge dV_{p+1}(adS)
\ee
where $dV_{p+1}(adS)$ is the volume $(p+1)$-form of the unit radius adS space. We may choose a gauge  for which the electric $(p+1)$-form field is
\be
C_{p+1} = ER^{p+2} f(y) dV_{p+1}(dS)\, , \qquad f'(y) = \cosh^{p+1}y\, . 
\ee
Using these results in (\ref{Dirac}) we find the effective action
\be
S[y] = -\mu R^{p+1}\int dV_{p+1}(adS)\left\{ \cosh^p y\sqrt{\cosh^2y + 
\bar g^{ij}_{adS}\partial_i y\partial_j y} -ER \, f(y)\right\}\, . 
\ee
Solutions of the equations of motion with constant $y$ are easily found by extremizing the effective potential 
\be
V_{eff}(y) = \mu R^{p+1} \left[ \cosh^{p+1}y -ER f(y)\right]\, . 
\ee
There is a minimum of this potential when
\be
\tanh y= \frac{ER}{p+1}\, , 
\ee
and in this case the induced worldvolume metric is adS with radius $R_{ind}$ given by
\be
\frac{1}{R_{ind}^2} =  \frac{1}{R^2} - \left(\frac{E}{p+1}\right)^2\, . 
\ee
This makes sense only if 
\be\label{boundE}
|E| <  E_{crit} \equiv \frac{p+1}{R}\, . 
\ee
Otherwise, the assumption that the worldvolume has adS geometry must be false. One can guess that 
if $|E|>E_{crit}$ the worldvolume geometry will be dS rather than adS, and we  verify this shortly. For the moment, we must restrict $E$ to be `subcritical'.  As the worldvolume is an adS space, its acceleration $A$ in the higher-dimensional flat space is such that 
\be
A^2 =- R_{ind}^{-2} = -R^{-2} + a^2 \, , \qquad a = |E|/(p+1)\, , 
\ee
where
\be
a < a_{crit}  \equiv E_{crit}/(p+1) = R^{-1}.
\ee
Because $A^2<0$, the brane temperature is zero.

\subsubsection{dS worldvolume}

We now consider a dS foliation of the adS spacetime by writing the adS metric in the form
\be
ds^2_{p+2} = R^2\left[ dy^2 + \sinh^2y \bar g^{dS}_{ij} d\sigma^i d\sigma^j \right]\, .
\ee
As before we identity the unit-radius dS coordinates $\sigma^i$ ($i=0,1,\dots,p$) with the worldvolume coordinates of the $p$-brane, in which case the induced worldvolume metric is 
\be
ds^2_{ind} = R^2 \left[ \sinh^2y \bar g^{dS}_{ij} + \partial_i y \partial_j y\right]\, d\sigma^id\sigma^j\, . 
\ee
As before, we assume a constant uniform electric-type field of strength $E$, so that
\be
F_{p+2} = ER^{p+2} \sinh^{p+1}y \, dy\wedge dV_{p+1}(dS)
\ee
where $dV_{p+1}(dS)$ is the volume $(p+1)$-form of the unit radius dS space. We may choose a 
gauge  for which the electric $(p+1)$-form field is
\be
C_{p+1} = ER^{p+2} f(y) dV_{p+1}(dS)\, , \qquad  f'(y) = \sinh^{p+1}y\, . 
\ee
Using these results in (\ref{Dirac}) we find the effective action
\be
S[y] = -\mu R^{p+1}\int dV_{p+1}(adS)\left\{ \sinh^p y\sqrt{\sinh^2y + 
\bar g^{ij}_{adS}\partial_i y\partial_j y} -ER \, f(y)\right\}\, . 
\ee
Solutions of the equations of motion with constant $y$ are easily found by extremizing the effective potential 
\be
V_{eff}(y) = \mu R^{p+1} \left[ \sinh^{p+1}y -ER f(y)\right]\, . 
\ee
There is a maximum of this potential when
\be
\coth y= \frac{ER}{p+1}\, , 
\ee
and in this case the induced worldvolume metric is dS with radius $R_{ind}$ such that 
\be
\frac{1}{R_{ind}^2} =  \left(\frac{E}{p+1}\right)^2 - R^{-2}\, . 
\ee
This makes sense only for $|E|>E_{crit}$, which we now assume because otherwise  the initial assumption of dS worldvolume geometry is false. Since the worldvolume geometry {\it is} dS, 
the brane acceleration $A$ is such that $A^2= R_{ind}^{-2}$, and hence the brane temperature is
given by
\be
2\pi T= \sqrt{a^2 - R^{-2}}\, ,\qquad a= |E|/(p+1) > R^{-1} \equiv a_{crit}\, . 
\ee

\subsubsection{Minkowski worldvolume}

One may guess from the above results that when $|E|=E_{crit}$ the worldvolume geometry will  be flat, and hence locally Minkowski. The simplest way to confirm this is to write the adS spacetime metric 
as
\be
ds^2_{p+2} = R^2 \left[ dy^2 + e^{2y} \eta_{ij} d\sigma^id\sigma^j \right]
\ee
where $\eta$ is the flat Minkowski metric, in local coordinates $\sigma^i$ ($i=0,1,\dots,p$). Again, we identify 
these coordinates with the brane worldvolume coordinates, in which case the induced metric is
\be
ds^2_{ind} = R^2\left[ e^{2y} \eta_{ij}  + \partial_i y\partial_j y\right] d\sigma^id\sigma^j\, . 
\ee
We assume a constant uniform electric-type field of strength $E$, so that
\be
F_{p+2} = ER^{p+2} e^{(p+1)y} \, dy\wedge d\xi^0 \wedge d\xi^1 \wedge \cdots \wedge d\xi^p\, , 
\ee
and we may then choose a gauge such that
\be
C_{p+1}= \left(\frac{ER}{p+1}\right) R^{p+1} e^{(p+1)y} d\sigma^0 \wedge d\sigma^1 \wedge \cdots \wedge d\sigma^p\, . 
\ee
Using these results in (\ref{Dirac}) we find the effective action
\be
S[y] = - \mu R^{p+1} \int d^{p+1}\xi\, e^{(p+1)y}\, \left\{ \sqrt{ 1+ e^{2y}\eta^{ij}\partial_i y \partial_j y} -  
\frac{ER}{p+1} \right\}\, .  
\ee

The effective potential for this case is
\be
V_{eff} =\frac{ \mu R^{p+1}}{p+1} e^{(p+1)y}\left[ E_{crit} - E \right] \, , 
\ee
where $E_{crit}= (p+1)/R$, as in (\ref{boundE}). This potential has no extrema, except when $E= E_{crit}$, in which case $V_{eff}\equiv 0$, so that there is a solution for any constant $y$.  For all these solutions the induced metric is locally Minkowski, but the brane acceleration is nevertheless non-zero. One can show that $a= R^{-1}$, the critical acceleration, as expected. If the same problem is analysed in 
terms of the brane moving in the ambient flat space of dimension $(p+3)$, with two time dimensions, in which the adS spacetime is embedded then one finds that the acceleration vector $A^\mu $ is null, and hence $A^2=0$ consistent with the zero brane temperature.

\section{BTZ black hole}

As a final application of our formalism, we consider the stationary  `BTZ'  metric representing a 
$(1+2)$-dimensional black hole  spacetime \cite{Banados:1992wn,Banados:1992gq}. 
This was found as a solution of  the $(1+2)$-dimensional Einstein's equations with negative cosmological constant  but it  can be globally embedded in a flat $(2+2)$-dimensional spacetime \cite{Deser:1998xb} and could therefore be interpreted as the worldvolume of a membrane, albeit with dynamics that differs from that assumed so far. Of course,  any metric that can be  globally embedded in a higher-dimensional flat spacetime can be viewed as a brane worldvolume  for  kinematical purposes.
The BTZ black hole  provides a simple illustration of this point, and it allows us to explore some issues arising for metrics that are stationary but not static. 

Also, the BTZ metric is locally diffeomorphic to $adS_3$. We have previously considered strings in $adS_3$ as a special case of $p$-branes in $adS_{p+2}$, but the global identifications of $adS_3$ needed to get the BTZ metric allows a further possibility for closed strings that we also examine. 

\subsection{Non-rotating black hole}

We begin with a brief discussion of the non-rotating BTZ black hole of mass $M$, for which the metric is 
\be
ds^2 = -\left({r^2\over R^2} - M\right) dt^2+ \left({r^2\over R^2} - M\right)^{-1} dr^2 + r^2d\phi^2\ ,
\ee
where $\phi$ is an angular coordinate with the standard $2\pi$  identification. This metric is static, with Killing vector field $k= \partial_t$. The singularity of the metric at $r=R \sqrt{M}$ is a coordinate singularity at a horizon of $k$, and the Euclidean continuation of the metric is non-singular if the imaginary time is identified with period $2\pi R/\sqrt{M}$, so the Hawking temperature  is 
$T_H = \sqrt{M}/(2\pi R)$. Actually, there is no absolute meaning to this temperature (in contrast to the Hawking temperature for asymptotically flat black holes) because $k^2\to 0$ as $r\to\infty$ and there is therefore no natural normalization for $k$. However, a rescaling of $k$ leads to a rescaling of $T_H$ such that the local temperature, satisfying the Tolman law (\ref{Tolman}) is unchanged. This local temperature is
\be\label{localBTZ}
T= \frac{T_H}{\sqrt{-k^2}} = \frac{1}{2\pi }\sqrt{\frac{M}{r^2- MR^2} }\, . 
\ee
This result can be reproduced \cite{Deser:1998xb} by applying the Unruh formula to the global embedding  of the BTZ  metric in a flat spacetime, which we give  for the general stationary BTZ black hole in the following subsection. A shortcut is to observe that since the BTZ metric is locally  diffeomorphic to $adS_3$, the temperature is given by the formula (\ref{DL}), with $a$ the acceleration of an observer in the BTZ metric at fixed $(r,\phi)$, which is
\be
a= \frac{r}{R\sqrt{r^2-MR^2}}\, . 
\ee
Using this in (\ref{DL}) we recover  (\ref{localBTZ}).  

Now consider a circular string  in the  BTZ  background,  at constant  $r$. This implies that the string
is subject to a force that prevents it from shrinking towards the BTZ horizon  but  just such a force results from a minimal coupling of the string to an appropriate constant uniform  electric-type field; we omit the details as they are given in more generality below.  The induced worldsheet metric for this circular-string solution is flat  but the extrinsic curvature is non-zero. Computing the extrinsic curvature for the circular string embedded in the four-dimensional flat ambient space
one gets (see also below)
\be
A=K_{uu} = \sqrt{\frac{M}{r^2- MR^2} }\, . 
\label{keepit}
\ee
from which we deduce that the string temperature equals the BTZ black hole temperature at $r=r_0$, and hence that the string is in thermal equilibrium with its background. 

\subsection{Rotating black hole}

We now turn to the case of the rotating BTZ black hole, which introduces a number of novel features. 
The metric takes the form
\be\label{BTZmetric}
ds^2 = -N^2 dt^2+ N^{-2} dr^2 + r^2\left(d\phi +N^\phi dt\right)^2\ , 
\ee
where 
\bea
N^2 &=& {1\over r^2R^2}(r^2-r_+^2)(r^2-r_-^2)={r^2\over R^2} - M +{J^2\over 4 r^2}\ ,\nn
N^\phi &=& -{r_+r_- \over r^2 R} = -{J\over 2 r^2}\, .  
\eea
for constants $r_\pm$, in terms of which  the mass and angular momentum are, respectively,
\be
M ={r_+^2+r_-^2\over R^2} \ ,\qquad J={2r_+ r_-\over R}\, . 
\ee
This metric has coordinate singularities at $r=r_+$ and $r=r_-$. Observers at fixed $r$ that follow 
orbits of the vector field
\be\label{ZAMO}
\xi = \partial_t - N^\phi(r) \partial_\phi\, . 
\ee
are known as ZAMOs, or zero angular momentum observers. The vector field $\xi$ is 
not Killing, but the vector fields
\be
\xi_\pm = \xi|_{r= r_\pm}
\ee
{\it are} Killing, and the coordinate singularity  at $r=r_\pm$ is a  Killing horizon of $\xi_\pm$.
In particular
\be
\xi_+^2 = -\frac{\left(r_+^2-r_-^2\right)\left(r^2-r_+^2\right)}{R^2r_+^2}\, , 
\ee
which shows that $\xi_+$ is timelike for $r>r_+$ and null at $r=r_+$.  Thermal equilibrium for $r>r_+$ requires that $T= T_H/\sqrt{-\xi_+^2}$, where $T_H$ is the Hawking temperature, as determined by requiring non-singularity of the Euclidean metric (with imaginary $J$): 
\be
T_H= \frac{r_+^2-r_-^2}{2\pi\, r_+R^2}\, . 
\ee
This gives the local temperature
\be\label{localTbtz}
T= \frac{1}{2\pi R} \sqrt{\frac{r_+^2-r_-^2}{r^2-r_+^2}}\, . 
\ee

The BTZ  metric can be globally embeded in a flat  $(2+2)$-dimensional spacetime  with metric
\be
ds^2 =-(dX_0)^2+(dX_1)^2+ (dX_2)^2 -(dX_3)^2
\ee
The embedding is  \cite{Deser:1998xb}
\bea
X_0 \pm X_1 &=& \pm R\sqrt{r^2-r_+^2\over r_+^2 -r_-^2} 
\exp\left[\pm\left({r_+ t\over R^2} -{r_-\phi \over R}\right)\right]\nn
X_2\pm X_3 &=&  \pm R\sqrt{r^2-r_-^2\over r_+^2 -r_-^2} 
\exp\left[\pm\left({r_+ \phi\over R} -{r_-t\over R^2}\right)\right]\, . 
\eea
In other words, the BTZ metric is the induced  metric on the hypersurface specified by this embedding, 
If we view this hypersurface as the worldvolume of a membrane then we may apply the formalism developed in the previous examples. In particular, we read off from the induced metric that the 
worldvolume velocity is
\be
u=N^{-1} \xi
\ee
where $\xi$ is the vector field of (\ref{ZAMO}). Thus, our definition of brane acceleration picks out ZAMOs as `preferred' observers.  

It is now a straightforward excercise to compute the brane velocity $U$, brane acceleration $A$ and brane jerk $\Sigma$. One finds that
\bea
U_0 \pm U_1 &=& \frac{r_+}{r}\sqrt{\frac{r^2-r_-^2}{r_+^2-r_-^2}}
\exp\left[\pm\left({r_+ t\over R^2} -{r_-\phi \over R}\right)\right]\nn
U_2\pm U_3 &=& -\frac{r_-}{r}\sqrt{\frac{r^2-r_+^2}{r_+^2-r_-^2}}
\exp\left[\pm\left({r_+ \phi\over R} -{r_-t\over R^2}\right)\right]\, . 
\eea
and
\bea
A_0\pm A_1 &=& \pm \frac{r_+}{Rr}\sqrt{\frac{r^2-r_-^2}{r^2-r_+^2}} \left(U_0\pm U_1\right)\nn
A_2\pm A_3 &=& \mp \frac{r_-}{Rr}\sqrt{\frac{r^2-r_+^2}{r^2-r_-^2}} \left(U_2\pm U_3\right)\ . 
\eea
One may verify that $U^2=-1$ and $U\cdot A=0$, and a simple computation shows that
\be\label{Asquared2}
A^2= \frac{r^6\left(r_+^2+r_-^2\right)-3r^4r_+^2r_-^2 + r_+^4r_-^4}
{r^4R^2\left(r^2-r_+^2\right)\left(r^2-r_-^2\right)}\, . 
\ee
Similarly, one can show that
\bea
\Sigma_0 \pm \Sigma_1 &=& -\frac{r_-^2\left(r^4-r_+^2r_-^2\right)}{R^2r^4\left(r^2-r_-^2\right)}
\left(U_0\pm U_1\right)\nn
\Sigma_2 \pm \Sigma_3 &=& -\frac{r_+^2\left(r^4-r_+^2r_-^2\right)}{R^2r^4\left(r^2-r_+^2\right)}
\left(U_2\pm U_3\right)\, , 
\eea
and hence that
\be
\lambda \equiv  \sqrt{\Sigma^2}/A^2 = 
\frac{r_-r_+\left(r^4-r_+^2r_-^2\right)\sqrt{\left(r^2-r_+^2\right)\left(r^2-r_-^2\right)}}
{r^6\left(r_+^2+r_-^2\right) -3r^4r_+^2r_-^2 + r_+^4r_-^4}
\label{lovv}
\ee
We recover the non-rotating case by setting $r_-=0$. In this case $\lambda=0$ and the Unruh formula applies.  In addition, $\lambda \to 0$ as  $r\to r_+$, so the Unruh formula applies near the horizon.
Comparison with (\ref{localTbtz}) shows that
\be
|A| = 2\pi T \left[1+ {\cal O}\left(\sqrt{r-r_+}\right)\right]\, , 
\ee
so the Unruh formula also gives the temperature predicted by thermal equilibrium, near the horizon. 
This is entirely as expected because the ZAMOs follow orbits of $\xi_+$ near the horizon only, and 
away from the horizon,  the Unruh formula is not applicable. 

\subsubsection{Jerk-free observers}

We have seen that the brane acceleration, as defined by the choice of coordinates in which the BTZ metric is given by (\ref{BTZmetric}), does not reproduce the local temperature (\ref{localTbtz}), except near the horizon, for understandable reasons: the Unruh formula is not valid away from the horizon because the parameter $\lambda$ is small only near the horizon.  However, 
let us consider 
 local observers that follow orbits of the Killing vector field 
\be
u_+ = \xi_+/\sqrt{-\xi_+^2}\, . 
\ee
These observers are circling the black hole with the angular velocity of the horizon.
As noticed in \cite{Deser:1998xb}, they experience a temperature equal to the local equilibrium temperature  (\ref{localTbtz}).
Following our formalism and taking $u=u_+$, we define 
\be
U_+ = u_+^i\partial_i X\, , \qquad A_+= u_+^i\partial_i U\, , \qquad
\Sigma_+ = u_+^i\partial_i A - A^2 U_+\, . 
\ee
A calculation shows that 
\be
A_+^2 = \frac{r_+^2-r_-^2}{R^2\left(r^2-r_+^2\right)}\, , \qquad \Sigma_+\equiv 0\, . 
\ee
It follows that the parameter $\lambda$ is zero, so the Unruh formula is applicable; 
it yields precisely the local temperature  (\ref{localTbtz}). This result is in agreement with a general principle
discussed in  \cite{Brown:1990fk} that a heat bath in equilibrium with a rotating black hole must itself rotate with the angular velocity of the horizon.

\subsection{Circular string}

Now we consider a circular string at constant $r$. We may identify $(t,\phi)$ with the worldsheet coordinates.  The induced metric is then
\be
ds^2_{ind}= -{1\over R^2}(r^2-r_+^2-r_-^2) dt^2-2{r_+r_-\over R}  d\phi dt +r^2d\phi^2
\ee
Using this in the Nambu-Goto action, and allowing for a uniform electric-type  2-form field, as
in previous examples, one finds a string action from which we obtain the following 
effective potential:
\be
V_{\rm eff} = \mu \ r \left[ \sqrt{ {r^2\over R^2} - M +{J^2\over 4 r^2}} - {e r\over R}  \right]\, ,
\qquad e\equiv {1\over 2}|E|R\ , 
\ee
There is an $e$, with $e^2>1$,  such that this potential has a maximum at any $r=r_0> r_+$, so we may adjust $e$ to get a static string solution at any distance from the black hole horizon. 
We now apply our previous formalism to find a brane temperature. Firstly, the worldsheet velocity
is $u= N^{-1}\xi$, exactly as it was for the BTZ black hole itself,  except that $\xi$ is now a worldsheet 
Killing vector field bcause $r=r_0$. This means that the calculation of $U,A,\Sigma$ for the string is {\it identical} to the calculation just done for the BTZ black hole. In particular,  $A_{string}^2$ is given 
by (\ref{Asquared2}) with $r=r_0$.
Alternatively, we can compute $A^2$ of the string from the extrinsic curvature of the worldsheet in the 
Minkowski spacetime, because a simple computation shows that the worldsheet acceleration is zero. 
The non-zero components of the extrinsic curvature tensor are 
\bea
K_{uu}^0 \pm K_{uu}^1 &=& \pm \frac{r_+^2\left(r_0^2-r_-^2\right)}
{r_0^2R\sqrt{\left(r_0^2-r_+^2\right)\left(r_0^2-r_-^2\right)}} 
\exp\left[\pm \left(\frac{r_+t}{R^2}-\frac{r_-\phi}{R}\right)\right]\, , \nonumber\\
K_{uu}^3 \pm K_{uu}^4 &=& \pm \frac{r_-^2\left(r_0^2-r_+^2\right)}
{r_0^2R\sqrt{\left(r_0^2-r_+^2\right)\left(r_0^2-r_-^2\right)}} 
\exp\left[\pm \left(\frac{r_+\phi}{R}-\frac{r_-t}{R^2}\right)\right]\, . 
\eea
We may now use $A^2 = K_{uu}^2$ to verify that  $A_{string}^2 $ is given 
by (\ref{Asquared2}) with $r=r_0$. 

Exactly the same calculation that led to (\ref{lovv}) now gives  $\lambda_{string}$ on setting $r=r_0$.
Therefore the Unruh formula is applicable only near the horizon ($r_0\to r_+$) where it
gives a temperature approximately equal to the local temperature (\ref{localTbtz}) of the black hole.

\section{Discussion}

It is a remarkable fact, established in recent years, that the local temperature of a stationary black hole in thermal equilibrium is  the local Unruh temperature associated to acceleration  in a flat ambient spacetime in which the black hole spacetime is globally and isometrically embedded.  As this
interpretation of black hole temperature relies only on the {\it kinematics} of global embeddings, it is equally applicable to relativistic branes,  for which the dynamics arises not from Einstein's equations 
`on the brane'  but rather  from the forces due to brane tension and minimal coupling to a background 
$(p+1)$-form potential.   In this paper we have shown that the Hamiltonian formulation of brane dynamics  leads naturally  to a `brane velocity'  in the ambient spacetime, and a `brane acceleration'  to which the Unruh formula may be applied to deduce a `brane temperature'.

Although the Unruh effect  was originally derived for a particle undergoing constant  proper 
acceleration, it is obvious on physical grounds that the result must continue to apply to a good approximation when the acceleration is allowed to vary sufficiently slowly in time. In the context of 
brane acceleration, which reduces for $p=1$ to the standard relativistic  acceleration, we 
constructed a dimensionless quantity $\lambda$ from the acceleration and a relativistic `brane jerk'
that provides a measure of the time rate of change of acceleration. When $\lambda\ll 1$ we expect the Unruh formula to be applicable, but otherwise it will not apply. 
Naturally, our definition of brane  jerk  applies, as a special case, to the standard relativistic mechanics of a particle in $(1+3)$-dimensional Minkowski spacetime with 4-velocity $u$ and $4$-acceleration $a$, and in that case the 4-vector $j= da/d\tau$ (which might be considered the natural relativistic generalization of $3$-jerk)  has the curious feature that it does not vanish even when the 3-acceleration is constant; the  4-vector that measures the rate of change of the 3-acceleration is $\varsigma= j-a^2 u$. We are not aware of any standard name for this quantity, but it would make sense to call this the relativistic jerk. In any case, a similarly defined quantity $\Sigma $ for brane jerk was needed in our definition of the parameter $\lambda$. 

As our first example, we considered an open $p$-brane undergoing rigid rotation in a $(2p+1)$-dimensional Minkowski spacetime, in the absence of any external forces. We remark here that the 
same solution applies to a `folded' closed $p$-brane in  which the $(p-1)$-dimensional `fold' moves 
at the speed of light, just like the boundary of an open $p$-brane. For $p=5$, this yields what appears to be a new solution of the M5-brane equations in the 11-dimensional Minkowski background of  M-theory.  It is convenient to describe this solution in a co-rotating frame in which the obvious timelike Killing vector field has a $p$-dimensional horizon. This Killing horizon of the Minkowski spacetime becomes a curvature singularity of the induced worldvolume metric, albeit one with a well-understood physical interpretation: it is the singularity on  the null hypersurface swept out by the $(p-1)$-dimensional boundary of the open $p$-brane. It is worth noting  that this provides an example of a curvature singularity that needs no   `resolution'  in the context of some more general theory. No doubt this case is exceptional, but it shows that the usual assumption that curvature singularities {\it require} new physics  is not self-evident.

In this example, the parameter $\lambda$ is always greater than unity, so we cannot expect 
to associate the centripetal acceleration with a temperature. Indeed, the attempt to do so leads to a 
local temperature that is not consistent with thermal equilibrium.  This is probably  related to the fact that the singularity of the worldvolume metric at the $p$-brane boundary is a curvature singularity rather than a coordinate singularity.  It is instructive to compare the worldvolume metric of the
rigidly-rotating $p$-brane with that of $(p+1)$-dimensional de Sitter spacetime in static coordinates. 
Consider $p=1$ for simplicity. Then, by rescaling coordinates,  both metrics can be put in the form
\be
ds^2 = -\left(1-\rho^2\right) dt^2 + f(\rho) d\rho^2 
\ee
for a function $f$, with $f=1$ for the rotating string and $f= 1/(1-\rho^2)$ for $adS_2$. Both are singular at $\rho=1$,  and in both cases the local acceleration in the embedding Minkowski spacetime becomes infinite as this singularity is approached. However, in the de Sitter case the singularity is a 
coordinate singularity  and the local acceleration is never zero; it takes a minimum at $\rho=0$, and the Unruh  temperature there is the Gibbons-Hawking temperature. In contrast, the singularity in the rotating brane case is a curvature singularity, and the local acceleration is zero at $\rho=0$. 

For the $p=1$ case, our rigidly-rotating brane example is closely related to the motion of particles in accelerators. A `circular Unruh'  effect was considered long ago in this context  \cite{Bell:1986ir} (see \cite{Akhmedov:2007xu} for an up-to-date account). As already noted in \cite{Bell:1986ir},   the `circular Unruh'  effect  is not thermal, and so cannot be  simply characterized by a temperature. This is presumably related to the fact that one cannot ignore the jerk of a particle undergoing circular motion. 

Most of our other examples involved spherical $p$-branes coupled to a constant and uniform electric-type background field. The simplest of these involve a $p$-brane in a $(p+2)$-dimensional spacetime. 
There is a static solution, at an unstable equilibrium point of the `effective' potential of the $p$-brane action. As expected, this has zero brane acceleration  and hence zero brane temperature. 
A slight increase of the $p$-brane radius leads to a runaway expanding brane solution; such solutions were considered previously in the context of a  brane generalization of the Schwinger pair-creaton process for charged particles in $(1+1)$-dimensional electrodynamics \cite{Dowker:1995sg}. A new observation of this paper is that the asymptotic expanding-brane solution has de Sitter worldvolume geometry. We also show that the brane acceleration of this asymptotic solution is constant and uniform, and that the associated Unruh temperature  is the Gibbons-Hawking temperature of de Sitter space. 

There is a close similarity of these results to cosmology with the force due to tension playing the role of gravitational attraction and the electric-type field playing the role of a positive cosmological constant . The static brane solution is both analogous and isometric to the Einstein static universe solution of Einstein's equations, and the de Sitter brane solution  is both analogous and isometric to the de Sitter solution of Einstein's equations.  Recall that the Einstein static universe is unstable, and that there exists a perturbation of that leads to an expanding universe that approaches the de Sitter universe at late times; this interpolating solution of Einstein's equations is analogous to  the solution of the brane equations that interpolates between the static and de Sitter branes. This analogy suggests 
that the observed small cosmological constant  might be explained by a bulk electric field in 
some braneworld scenario; if so,  the question of why the cosmological constant is small could be 
traded for a similar question for the bulk electric field, but  the latter question might have some simple answer (e.g. because electric field is produced by vacuum fluctuations).  It would be  interesting to find a concrete scenario in which this possibility could be explored. 

We have also studied spherical $p$-branes in $(p+2)$-dimensional de Sitter and anti-de Sitter spacetimes. Since these may be embedded in a flat $(p+3)$-dimensional spacetime, albeit with two time dimensions in the adS case, this problem is closely related to one in which a $p$-brane is embedded in a flat $(p+3)$-dimensional spacetime, so 
that the worldvolume has co-dimension two in the flat ambient spacetime. Viewed from this perspective, the brane acceleration $A$, which is orthogonal to the 
worldvolume, now has two components: one is due to the acceleration $a$ of the brane in (a)dS and the other to the acceleration of (a)dS in the flat spacetime. This yields $A^2= a^2 \pm R^{-2}$ for an (a)dS spacetime of radius $R$, where the plus sign is for de Sitter and the minus sign for  anti-de Sitter. 
The adS case is more interesting because  a non-zero temperature requires  $a>a_{crit}$ for  `critical' acceleration $a_{crit}= 1/R$. We have shown that the brane worldvolume is  a $(p+1)$-dimensional adS spacetime for  $a<a_{crit}$ whereas it is a $(p+1)$-dimensional  dS spacetime for $a>a_{crit}$, with a local temperature that again equals the Gibbons-Hawking temperature. 

It is interesting to ask what significance the de Sitter embeddings in adS might have for the adS/CFT correspondence. Recall that the $adS_{p+2}$ spacetime has a boundary that is topologically 
$S^1\times S^p$. An expanding bulk $p$-brane with de Sitter worldvolume could be the end result of a instanton-induced creation process that leads to an expanding bubble in the bulk that, asymptotically, 
wraps the $p$-sphere at infinity. It is natural to suppose that the asymptotic approach of the brane temperature to the Gibbons-Hawking temperature of a dS spacetime corresponds to some approach to thermal equilibrium on the boundary, and this implies that the boundary temperature is non-zero. 

In our  final example we reconsidered the BTZ black hole, studied from the GEMS perspective in 
\cite{Deser:1998xb,Hong:2000kn}. As we focus on the kinematics, we may suppose that it is a membrane in $(1+3)$-dimensional Minkowski spacetime, and then ask what its brane acceleration is according to our formalism. We found results that are close to those found in 
\cite{Deser:1998xb}
 because  our definitions effectively select the class of observers known as ZAMOs, which are also those considered in \cite{Brown:1990fk} and in earlier discussions of BTZ black holes \cite{Carlip:1995qv}. 

As a final comment, we recall  that the Minkowski metric in the form (\ref{dSsliceMink}), foliated by de Sitter hypersufaces,  is the natural higher-dimensional generalization of the $(1+1)$-dimensional Rindler spacetime;  the analog of the Unruh temperature in Rindler space is the Gibbons-Hawking temperature on  a de Sitter slice.  This suggests a  new kind of holography because the isometry group of the  Rindler-type wedge is $SO(p+1,1)$, which is the same as the isometry group of $dS_{p+1}$.
Thus, gravitational physics on the Rindler-type wedge with a cut-off at $r=R$ could have a holographic description in terms of a quantum field theory on the de Sitter slice at $r=R$. 
One may also wonder whether higher-dimensional Rindler spacetimes can 
arise as near-horizon geometries in the same way that the original, 
$(1+1)$-dimensional, Rindler spacetime arises in the context of the 
Schwarzschild black hole.

\bigskip
\noindent
\section*{Acknowledgments}
We thank  Roberto Emparan, Gary Gibbons, Cesar Gomez  and Robert Myers for helpful discussions. We are also grateful to Roberto Iengo for comments on, 
and corrections to, the first version of this paper.
JGR acknowledges support by MCYT FPA 2007-66665, European
EC-RTN network MRTN-CT-2004-005104 and CIRIT GC 2005SGR-00564. PKT is supported by
an ESPRC Senior Research Fellowship.

\setcounter{section}{0}

\end{document}